\documentclass[hyper(*)]{JHEP3}

\usepackage{epsfig,multicol,bbm}
\usepackage{enumerate}
\usepackage{bibentry}
\usepackage{amsmath}
\title{Three dimensional Seiberg-like duality and tropical cluster algebra}

\author{Dan Xie

\\ School of Natural Sciences, Institute for Advanced Study \\
Princeton, NJ 08540, USA}

\abstract{Seiberg-like duality of three dimensional $\mathcal{N}=2$ Chern-Simons-Matter quiver gauge theory is shown to have a chiral double tropical cluster algebra structure. 
We use cluster algebra results to study combinatorial aspects of these theories such as classification, supersymmetry breaking, etc.}

\begin{document}
\maketitle

\section{Introduction}
Quiver gauge theories  built from bifundamental matter are a nice class of quantum field theories which can be used to learn interesting gauge dynamics. In particular,  IR dualities such
as Seiberg duality \cite{Seiberg:1994pq} are very rich for this class of theories.  Consider a four dimensional $\mathcal{N}=1$ quiver gauge theory $Q$, Seiberg dual theory  of $Q$  can be found by doing a sequence of Seiberg dualities  on various quiver nodes \cite{Berenstein:2002fi}. So local aspects of dualities  are well understood. The interesting question is global or combinatorial aspects of dualities, and some of them are:
 \begin{itemize}
 \item Find a method to decide whether two different looking quivers can be related by Seiberg duality.
 \item Find  true degree of freedoms of gauge theory, since some of the gauge groups might confine which will be manifest only in certain duality frame.
 \item Determine whether the supersymmetry is broken or not, since the rank of gauge group might become negative in certain duality frame.
 \end{itemize}
 These questions are typically not easy to answer even for very simple quiver gauge theory, as generically there are infinite number of duality frames! 
 
Fortunately, the combinatorial aspects of Seiberg duality is equivalent to cluster algebra \cite{Fomin2001}. The basic data of cluster algebra is also a quiver $Q$, and different 
quivers are related by so-called local cluster transformation acting on a single quiver node, whose action on quiver is exactly the same as Seiberg duality,
so one can use many remarkable results from cluster algebra to study combinatorial aspects of Seiberg dualities. For instance, the above three 
questions can be answered for an interesting class of  four dimensional quiver gauge theory based on planar network \cite{Xie:2012mr,Franco:2012mm}.

In this paper, we would like to study IR dualities of three dimensional $\mathcal{N}=2$ Chern-Simons matter (CSM) quiver gauge theory built from bifundamental matter. For three dimensional theory, there 
are some new features: first there are no chiral anomaly cancellation  and  asymptotical free conditions on gauge group, so the assignment of gauge groups can be arbitrary and we 
can perform duality on any quiver node; second, one can
add Chern-Simons(CS) terms and this introduces a new type of discrete UV data besides gauge group ranks.

It was found by Aharony \cite{Aharony:1997gp} that a very similar Seiberg duality exists for three dimensional $\mathcal{N}=2$ SQCD with U(N) group (recently Seiberg duality 
of SU(N) group is found in \cite{Aharony:2013dha}), and the duality action on quiver is almost the same as the four dimensional Seiberg duality. 
 Later such Seiberg-like duality is generalized to theory with Chern-Simons term by Giveon and Kutasov \cite{Giveon:2008zn}. 
 The general duality formula for U(N) theory with arbitrary CS levels and chiral matter is presented in \cite{Benini:2011mf} \footnote{See \cite{Amariti:2009rb, Cremonesi:2010ae} for early attempt in finding Seiberg duality of 
 general CSM theory.}. 

So local aspects of dualities of 3d $\mathcal{N}=2$ CSM  quiver gauge theory is also understood. 
We would still like to ask the same global combinatorial questions about dualities, and it is natural to wonder whether three dimensional 
Seiberg duality is also related to cluster algebra. However, the formula found in \cite{Benini:2011mf} seems to imply that  such connection is not here.

The purpose of this paper is to show that  Seiberg-like duality for three dimensional quiver gauge theory indeed has a 
cluster algebra structure. To achieve this, we need to slightly modify our quiver $Q$: we need to add an extra quiver node to $Q$ and get an extended quiver $\tilde{Q}$, and  the arrows between this extra node and the original quiver nodes 
are determined by the Chern-Simons levels. With this realization, the formula for Seiberg-duality on a single gauge group is greatly simplified, and 
it is just cluster transformation on the extended quiver $\tilde{Q}$. 

In fact, the connection between three dimensional theory and cluster algebra is more 
satisfactory than four dimensional theory, since cluster algebra actually involves two sets of variables, which are naturally identified with the gauge groups and 
the CS levels.  

This paper is organized as follows: In section 2, we review basic elements of tropical cluster algebra; in section 3,
we show how  Seiberg-like duality of 3d $\mathcal{N}=2$ CSM theory can be interpreted as 
the cluster transformation on a slightly extended quiver; In section 4, we study  global combinatorial questions about the dualities; finally, we give a short conclusion.

\section{Introduction to tropical cluster algebra} 
Cluster algebra \cite{Fomin2001} is found to play a crucial role in various physical contexts, such as 
BPS counting and line operators \cite{Xie:2012gd, Gaiotto:2010be, Alim:2011kw, Xie:2013lca} of 4d $\mathcal{N}=2$ theory, scattering amplitude of 4d $\mathcal{N}=4$ theory \cite{arkani2012scattering}, and three dimensional 
$\mathcal{N}=2$ abelian CSM theory \cite{Dimofte:2013iv,Terashima:2013fg}.

Cluster algebra consists of a family of quadruples (called seeds)\footnote{Here we use the tropical version, see \cite{fock2005dual}; and the quiver is taken to be 2-cyclic: there are no one and  two cycles.} $(\text{Q}, \text{W}, a_i, x_i)$:
$\text{Q}$ is a quiver represented by an antisymmetric tensor $\epsilon_{ij}$;
$\text{W}$ is a potential associated with oriented cyclic paths in quiver;
$a_i$ and $x_i$ are a set of numbers \footnote{They can take rational, real or integer values depending on  interest.} defined on each 
quiver node. 

\begin{figure}[htbp]
\small
\centering
\includegraphics[width=6cm]{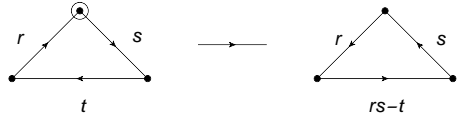}
\caption{The quiver mutation action on quiver node $k$: reverse the quiver arrows attached on quiver node $k$, and add a new quiver arrow between node $i$ and $j$ if there is an 
oriented path $i\rightarrow k \rightarrow j$, finally cancel two cycles between the quiver nodes. Here we assume $rs>t$. }
\label{seiberg1}
\end{figure}

These seeds are related by the cluster transformation defined on a quiver node $k$  \footnote{  
Quiver nodes are separated into two sets $(\text{I}, \text{I}_0)$: nodes in $\text{I}$ are dynamical which can be mutated,
 while nodes in $\text{I}_0$ are frozen and never mutated.}, and  new seed $(Q^{'}, W^{'}, a_i^{'}, x_i^{'})$ is related to the original one in a simple way.
The quiver changes as (also called quiver mutation):
\begin{equation}
\epsilon^{'}_{ij}=\left\{
\begin{array}{c l}
    -\epsilon_{ij}& \text{if}~i=k~\text{or}~j=k\\
    \epsilon_{ij}+\text{sgn}(\epsilon_{ik})[\epsilon_{ik}\epsilon_{kj}]_+ & \text{otherwise}
\end{array}\right.
\label{quiver}
\end{equation}
Here $\text{sgn}(\epsilon_{ik})=1$ if $\epsilon_{ik}>0$, and $[x]_+=max(x,0)$.
The graphical rule is: first change the orientation of quiver arrows attached to node $k$; then add  a new quiver arrow from $i$ to $j$ if there is an oriented path $i\rightarrow k \rightarrow j$,  
; finally, remove all the two cycles,
see figure. \ref{seiberg1}. 

\begin{figure}[htbp]
\small
\centering
\includegraphics[width=10cm]{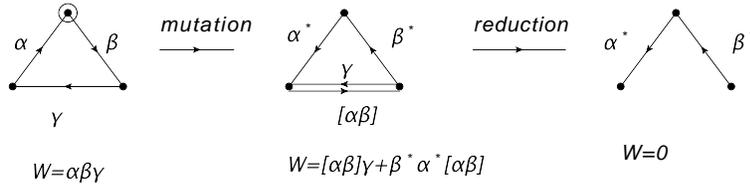}
\caption{The mutation action on the potential, and we get a reduced quiver and potential at the end. }
\label{seiberg}
\end{figure}

For the change of potential, consider an oriented path $\ldots i\xrightarrow{\alpha} k \xrightarrow{\beta} j\ldots$ passing through
node $k$ and assume the potential involving this piece has the following form
 \begin{equation}
 W=\sum \ldots\alpha\beta\ldots+\ldots;
 \end{equation} 
After mutation,   the orientations of $\alpha$ and $\beta$ are reversed (they are labeled by  $\alpha^{*}$ and $\beta^{*}$), and there is 
 a new  field $[\alpha\beta]$ between node $i$ and $j$. The potential changes in the following way:
  the  $\alpha\beta$ term in the original potential is replaced by the new field $[\alpha\beta]$, and there is an extra cubic potential
  term:
   \begin{equation}
 W^{'}=\sum\ldots[\alpha\beta]\ldots+\beta^{*}\alpha^{*}[\alpha\beta]+\ldots;
 \label{potential}
 \end{equation} 
Now there might be a quadratic term in $W^{'}$ involving field $[\alpha\beta]$, and one can integrate out fields appearing in quadratic potential. After reduction, we  get a reduced quiver with potential 
  $(Q_{\text{reduced}},W_{\text{reduced}})$\footnote{
We assume that the potential is a generic one such that one can always cancel the two cycles between two quiver nodes. This 
condition is implied in the mutation formula for the quiver.},
see figure. \ref{seiberg} for an example.

Finally, the cluster transformation formula on  tropical $a$ and $x$ variables are 
\begin{equation}
a^{'}_k= \text{max}(\sum_ia_i{[\epsilon_{ki}]_+},\sum_ia_i{[-\epsilon_{ki}]_+})-a_k 
\label{atrans}
\end{equation}
\begin{equation}
{x}_j^{'}=
\begin{cases}
    -{x}_k~~~~~~~~~~~~~~~~~~~~~~~~~~~~~~~~~~~~~~~j=k\\
   x_j+(\epsilon_{jk}) \text{max}(\text{sgn}(\epsilon_{jk})x_k,0)~~~~~~~j\neq k\\     
\end{cases}
\label{xtrans}
\end{equation}
There is an interesting duality between the $a_i$ and $x_i$ coordinates
\begin{equation}
x_i=-\sum_j\epsilon_{ij}a_j,
\label{dual}
\end{equation}
which can be verified by looking at the cluster transformation.

The above definition of  cluster algebra is closely related to Seiberg duality \cite{Seiberg:1994pq} if we regard 
$(Q,W)$ as defining a 4d quiver gauge theory with potential \footnote{For 4d theory, anomaly cancellation puts constraints on the rank of gauge group.} .
 In fact,  the rule for the change of the quiver  and potential  is exactly
the same as Seiberg duality \cite{Berenstein:2002fi}, and the example shown in figure. \ref{seiberg} is exactly the Seiberg duality  found in  \cite{Seiberg:1994pq}.
 Moreover, the  gauge group ranks can be identified with the $a$ variables as they have the same transformation rule under the duality. 
 For a 4d quiver gauge theory, the fundamental (outgoing arrows) and anti fundamental (incoming arrows) on a quiver node are the same, and let's denote the number of fundamentals on a quiver node as $N_f$, and 
 the transformation rule for the $a$ coordinates  [\ref{atrans}]  is 
 \begin{equation}
 N_i^{'}=N_f-N_i,
 \end{equation}
 which is exactly the transformation rule for gauge groups in Seiberg duality. 

Although the connection between 4d Seiberg duality and cluster algebra is quite interesting, there are several limitations:
 first chiral anomaly cancelation constraints the assignment of the gauge group, therefore the $a$ coordinates are constrained;
Secondly, the non-trivial Seiberg duality can only be done on quiver nodes in the conformal window, while the cluster transformation can be done on any node.
Thirdly, the $x$ coordinates do not enter into the story. 

The above constraints might be evaded if we regard the quiver as defining a 3d $\mathcal{N}=2$ theory, and consider 3d Seiberg-like duality \cite{Aharony:1997gp, Giveon:2008zn,Benini:2011mf}: firstly of all, 
the duality action on the quiver and potential is the same as 4d Seiberg duality \footnote{For three dimensional 
Seiberg-like dualities on gauge theory with adjoint matter, see \cite{Niarchos:2008jb, Kim:2013cma}. }; Secondly, there is no gauge anomaly and one can consider 
arbitrary gauge groups; and moreover there is no bound on gauge groups from asymptotical freedom and one can perform duality on any quiver node. Thirdly, there are extra UV Chern-Simons (CS) 
levels which might be identified as the tropical $x$ variable. 

However, the first glance on the duality formula found in \cite{Benini:2011mf} tells us 
3d Seiberg duality is not equivalent to the cluster transformation on tropical $a$ and $x$ coordinates of quiver $Q$ (see  \cite{Closset:2012eq} for the attempt to connect 3d Seiberg duality with quiver mutation).
In the following, we will show that three dimensional Seiberg-like duality is actually related to cluster algebra by slightly extending our quiver $Q$.

\section{3d $\mathcal{N}=2$ Seiberg duality and cluster algebra: local aspects}
Let's consider a 3d theory defined using a  2-acyclic quiver  and potential $(\text{Q},\text{W})$, here $\text{Q}$ is 
represented by an antisymmetric tensor $\epsilon_{ij}$, and quiver nodes are separated into two kinds: dynamical ones corresponding to gauge groups and frozen ones corresponding to flavor groups. We  assign gauge (flavor) group $\text{U}(\text{N}_i)$ on each quiver node, and add a bare CS term with level $k_i$ which satisfies the following condition:
\begin{equation}
k_i+{1\over 2}\sum_{j}\epsilon_{ij} N_j\in Z,
\label{anomy}
\end{equation}
which is due to parity anomaly cancellation. The UV data is shown in table. \ref{data1}, and many field theory aspects of three dimensional Chern-Simons matter theories can be found in \cite{Aharony:1997bx,Intriligator:2013lca}.
\begin{table}
\begin{center}
\begin{tabular}{|c|c|c|c| }
\hline                          
   UV data&Quiver and Potential & Gauge (flavor) group & CS level  \\ \hline
    Initial theory&$(\text{Q},\text{W})$&$\text{U}(\text{N}_i)$&$k_i$ \\ \hline
       Dual theory&$(\text{Q}^{'},\text{W}^{'})$:~[\ref{quiver}],[\ref{potential}]&$\text{U}(\text{N}_i^{'})$:~[\ref{rank}]&$k_i^{'}$:~[\ref{CS}] \\  \hline
\end{tabular}
\end{center}
  \caption{Summary of UV data defining initial  and dual theory.}
    \label{data1}
\end{table}

Let's use $(s_1^i, s_2^i)$ to denote the number of anti-fundamental (incoming arrows) and fundamentals (outgoing arrows) on a quiver node $i$:
\begin{equation}
s_1^i=\sum_j N_j [-\epsilon_{ij}]_{+},~~~s_2^i=\sum_j N_j [\epsilon_{ij}]_{+}, 
\end{equation}
and use  $k^{+}_i$ and $k^{-}_i$ to denote effective\footnote{One have two choices because the effective CS levels  depend on the sign of the expectation value of real scalar in the vector multiplet.}  CS level: 
\begin{equation}
k^{+}_i=k_i+{1\over 2}(s_2^i-s_1^i),~~k^{-}_i=k_i+{1\over 2}(s_1^i-s_2^i),
\end{equation}
and they are both integers due to [\ref{anomy}]. 

Under 3d Seiberg duality $\mu_i$ acting on a  gauge node $i$, the superpotential and quiver are changed as 4d Seiberg duality\footnote{A singlet and new potential 
term involving monopole operator is needed if there is a Coulomb branch.}, see [\ref{quiver}],[\ref{potential}], which is the same as the cluster transformation on quiver. 
The gauge group rank on node $i$ is changed as \cite{Benini:2011mf}:
\begin{equation}
N_i^{'}={1\over2}(s_1^i+s_2^i+|k^{-}_i|+|k^{+}_i|)-N_i,
\label{rank}
\end{equation}
and the  CS level is changed as:
\begin{equation}
k_j^{'}=\begin{cases}
-k_j~~~\text{if}~~~j=i\\
{1\over 2}(k_j^{+}+(-\epsilon_{ji})\text{max}(0,\text{sgn}(-\epsilon_{ji})k_i^{+}))+{1\over 2}(k_j^{-}+(\epsilon_{ji})\text{max}(0, \text{sgn}(\epsilon_{ji})k_i^{-})])~~~\text{if}~~~j\neq i\
\end{cases}
\label{CS}
\end{equation}
In the dual theory, there are also changes on CS levels between various $U(1)$ groups existing in our theories \cite{Benini:2011mf}, we do not consider those changes as those CS levels will not affect 
Seiberg duality on quivers and non-abelian groups.

\subsection{Chiral CS level and cluster algebra}
If the bare CS couplings of all the quiver nodes take the following values:
\begin{equation}
k_i={1\over 2}(s_1^i-s_2^i),
\end{equation}
then the two effective CS levels are
\begin{equation}
k_i^{+}=0,~~~~k_i^{-}=2k_i=s_1^i-s_2^i=-\sum_j\epsilon_{ij}N_j.
\end{equation}
We call  these CS levels as chiral CS levels. 
Then Seiberg duality actions on gauge group ranks and CS levels are simplified as: 
\begin{align}
&N_i^{'}=\text{max}(s_1^i,s_2^i)-N_i, \nonumber\\
&k_j^{-'}=\begin{cases}
&-k_j^{-}~~~\text{if}~~~j=i\\
&k_j^{-}+(\epsilon_{ji})\text{max}(0, \text{sgn}(\epsilon_{ji})k_i^{-})~~~~~j\neq i,
\end{cases}
\end{align}
and $k_i^{+}$ is still zero in dual theory. 
Comparing above formula with the cluster transformation shown in [\ref{atrans}] and [\ref{xtrans}], one can see that
the duality action on CS level $k_i^{-}$ and gauge group rank are the same as  $x$ and $a$ variables! Moreover, the nonzero effective CS level 
is related to the gauge group rank:
\begin{equation}
k^{-}_i=s_1^{i}-s_2^{i}=-\sum_j\epsilon_{ij}N_j,
\end{equation}
which is exactly the duality relation between two sets of cluster coordinates, see [\ref{dual}].

If we take the bare CS level at another chiral point $k_i={1\over 2}(-s_1^i+s_2^i)$, then  
\begin{equation}
k_i^{-}=0,~~~~k_i^{+}=-s_1^i+s_2^i=-\sum_j(-\epsilon_{ij})N_j.
\end{equation}
and the Seiberg duality action on CS level and gauge group is the same as the cluster transformation [\ref{atrans}] and [\ref{xtrans}] with the quiver 
replaced by chiral dual  $Q^o=-\epsilon_{ij}$:
\begin{align}
&N_i^{'}=\text{max}(s_1,s_2)-N_i, \nonumber\\
&k_j^{+'}=\begin{cases}
&-k_j^{+}~~~\text{if}~~~j=i\\
&k_j^{+}+((-\epsilon_{ji}))\text{max}(0, \text{sgn}((-\epsilon_{ji}))k_i^{+})~~~~~j\neq i.
\end{cases}
\end{align}
$k_i^{-}$ is zero in any duality frame, and we also have the duality relation between the CS level $k_i^{+}$ and the gauge group rank (using quiver $Q^o$.).
So the CS levels $k_i^{+}$ and the gauge group $N_i$ are identified with the tropical $x$ and $a$ coordinates of quiver  $Q^o$ at another chiral CS level.

\subsection{Generic CS level, extended quiver and chiral double}
For generic bare CS level $k_i$, the transformation rules on  CS levels factorize into two copies:
\begin{equation}
k_j^{-'}=\begin{cases}
-k_j^{-}~~~~~~~~~~~~~~~~~~~~~~~~~~~~~~~~~~~~~~~j=i\\
k_j^{-}+(\epsilon_{ji})\text{max}(0, \text{sgn}(\epsilon_{ji})k_i^{-})~~~j\neq i
\end{cases}
k_j^{+'}=\begin{cases}
-k_j^{+}~~~~~~~~~~~~~~~~~~~~~~~~~~~~~~~~~j=i\\
k_j^{+}+(-\epsilon_{ji})\text{max}(0, \text{sgn}(-\epsilon_{ji})k_i^{+})~~j\neq i
\end{cases}
\end{equation}
Notice that the initial data $k_i^{+}$ and $k_i^{-}$ are not independent, they
have to satisfy the constraint $k_i^{+}-k_i^{-}=s_2^i-s_1^i$. Comparing with the cluster transformation rule on tropical $x$ coordinates [\ref{xtrans}], we see that $k^{-}_i$ and $k^{+}_i$ can still
be identified as the tropical $x$ coordinates of $Q$ and $Q^o$.
 
Now let's look at the transformation rules on the rank of gauge groups.
Let's first rewrite  [\ref{rank}] in a different form:
\begin{equation}
N_i^{'}=
\begin{cases}
  \text{max}(k_i^{+}+s_1^i, s_2^i)-N_i,~~~ k_i^{+}\geq0 \\
    \text{max}(k_i^{+}+s_2^i, s_1^i)-N_i,~~~ k_i^{+}\leq0 \\
\end{cases}
\label{reformu1}
\end{equation}  
Since this formula mixes the CS levels (which is identified as the $x$ coordinates) and gauge group ranks, we can not regard gauge groups 
as the $a$ coordinates of quiver $Q$ or $Q^o$. 

\begin{center}
\begin{figure}[htbp]
\small
\centering
\includegraphics[width=10cm]{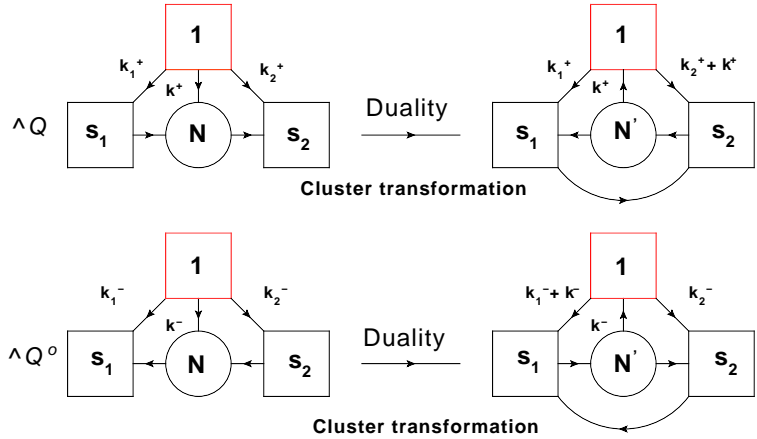}
\caption{3d Seiberg duality on a $U(N)$ gauge theory is realized as the cluster transformation on an extended quiver. }
\label{one}
\end{figure}
\end{center}

However, we recognize that the duality  transformation on gauge group can actually be regarded as the cluster transformation on 
an extended quiver $\tilde{Q}$ which is defined as follows: introducing a single extra quiver  node $f$ with rank one, and the quiver arrows between
$i$th quiver node to the extra node is 
\begin{equation}
\epsilon_{if}=-k_i^{+},
\end{equation}
See figure. \ref{one}.  Since the transformation of $x$ coordinates only depends on the quiver node under mutation,  the addition of 
new quiver nodes does not change the identification between the CS levels and the $x$ coordinates as the extra node is never
mutated.

Now let's look at the transformation rules for tropical $a$ coordinate of the extended quiver $\hat{Q}$, and
do mutation on node $i$ of $\hat{Q}$, the cluster transformation on $a$ coordinates of  $\hat{Q}$  becomes [\ref{atrans}]:
\begin{equation}
\tilde{N}_i^{'}=\text{max}(\tilde{s}_2,\tilde{s}_1)-N_i=\begin{cases}
  \text{max}(k_i^{+}+s_1^i, s_2^i)-N_i,~~~ k_i^{+}\geq0 \\
    \text{max}(k_i^{+}+s_2^i, s_1^i)-N_i,~~~ k_i^{+}\leq0 \\
\end{cases}
\label{Qd}
\end{equation}
and this formula is indeed the same as the duality transformation on the ranks of gauge group [\ref{reformu1}]. 

We would like to check that the new quiver $\tilde{Q}^{'}$ still satisfies our defining relation using the CS level. 
After doing the cluster transformation, all the arrows attached on quiver node $i$ is reversed, and in particular, we have
\begin{equation}
\epsilon_{if}^{'}=k_i^{+}.
\end{equation}
This is consistent since the new CS level is $k_i^{+'}=-k_i^{+}$, and to construct an extended quiver $\tilde{Q}^{'}$, we indeed
need to set $\epsilon_{if}^{'}=-k_i^{+'}=k_i^{+}$. 

Let's now check that $k_i^{-}$ and $N_i$ satisfy the duality formula on extended quiver $\tilde{Q}$ 
\begin{equation}
-\sum \tilde{\epsilon}_{ij} N_j=- \sum \epsilon_{ij}N_j-\epsilon_{if}=k_i^{+}+s_1^{i}-s_2^{i}=k_i^{-},
\end{equation}
so we have the following identification:
\begin{itemize}
\item $k_i^{-}$ can be identified with the tropical $x$ coordinates of $\tilde{Q}$.
\item $N_i$  can be identified with the tropical $a$ coordinates of $\tilde{Q}$.
\end{itemize}
And they satisfy the duality relation. Notice that $k_i^{+}$ is even easier to find, as they just equal to the number of arrows between the extra node and $i$th quiver node.  

Similarly, one can construct an extended quiver $\tilde{Q}^{o}$ using $k_i^{-}$. This can be seen if we rewrite the transformation rules on the gauge groups [\ref{rank}] as
\begin{equation}
N_i^{'}=
\begin{cases}
  \text{max}(k_i^{-}+s_2^i, s_1^i)-N_i,~~~ k_i^{-}\geq0 \\
    \text{max}(k_i^{-}+s_1^i, s_1^i)-N_i,~~~ k_i^{-}\leq0 \\
\end{cases}
\label{reformu}
\end{equation}  
The extended quiver $\tilde{Q}^{o}$ is constructed as follows: adding a new extra node with fixed rank one to $Q^o$. The new arrows are determined as 
\begin{equation}
\epsilon_{if}=-k_i^{-};
\end{equation}
The interested reader can check that the cluster transformation of $a$ coordinates on $\tilde{Q}^{o}$ indeed is the same as the transformation rule on rank of gauge groups.
With above observations, we have the following conclusion:

\textbf{Theorem}: The Seiberg duality on
 3d  $\mathcal{N}=2$ CSM quiver gauge theory has a chiral double cluster algebra structure specified by two extended quiver $\hat{Q}$ and $\hat{Q^o}$. 
 
It is interesting to note that such chiral double cluster algebra structure is already studied in \cite{Fock:math0311245}.

\newpage

\section{3d $\mathcal{N}=2$ Seiberg duality and cluster algebra: global aspects}
\subsection{Classification}
The cluster algebra structure of 3d Seiberg duality  has one very interesting application: one 
can use the result from cluster algebra literature to classify 3d SCFT. 
\subsubsection{Finite type: ADE quiver}
The first question one might be interested about the classification is: what is shape of quiver such that there are only finite number of duality frames? 
This question is identical to the question of classifying finite cluster algebra. The answer is found in cluster algebra literature \cite{fomin2003cluster}:

\textbf{Theorem}: The cluster algebra is finite if and only if the underlying quiver (the dynamical part) is mutation equivalent to ADE quiver!

Notice that the orientation of the ADE diagram is not important as they are in the same mutation class. This classification is independent of 
the assignment of the gauge group rank and the CS level, and it only depends on the shape of quiver! Moreover, this classification only depends on
dynamical nodes, and the extra node in $\tilde{Q}$ is frozen, so our modification of quiver will not affect the classification, as the dynamical part of $Q$ and 
$\tilde{Q}$ is the same.

Using this result, we are led to claim that three dimensional finite  CSM quiver gauge theory  has a ADE classification (the gauge group part).

\begin{center}
\begin{figure}[htbp]
\small
\centering
\includegraphics[width=14cm]{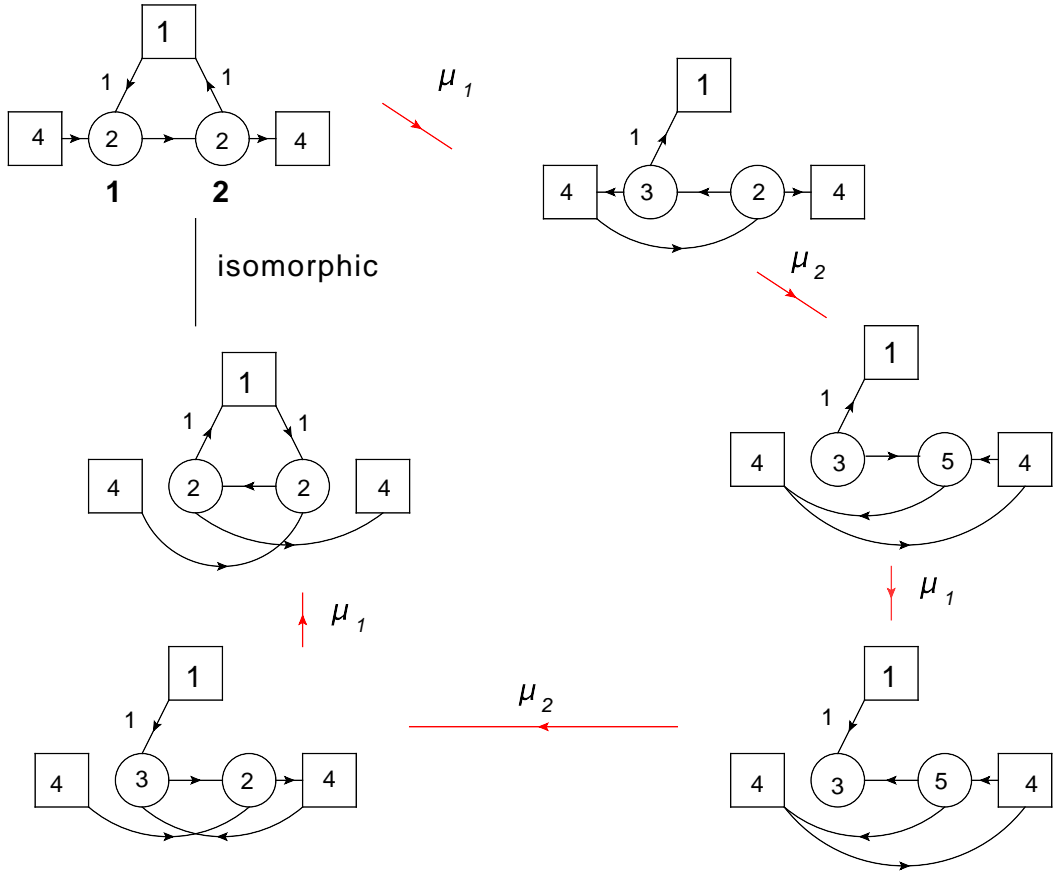}
\caption{The change of $A_2$ quiver under sequences of Seiberg duality, here we use extended quiver $\tilde{Q}$. }
\label{a2}
\end{figure}
\end{center}

\textbf{Example}: Let's consider a theory defined on $A_2$ Dynkin diagram. 
Let's start the quiver at the top of figure. \ref{a2}, and choose the rank of gauge group and flavor group as $(s_1, N_1, N_2, s_2)=(4,2,2,4)$. We 
take the bare CS level of two dynamical nodes as $(k_1, k_2)=(2, -2)$. Then we do a sequence of Seiberg duality  $(\mu_1, \mu_2, \mu_1, \mu_2, \mu_1)$ by starting mutating node one. The 
change of gauge group rank and the CS level is shown in table. \ref{data} (We do not show the change of CS level of the flavor nodes.), and the 
gauge theory come back to itself up to a permutation on quiver nodes!  For general ADE quiver gauge theory, One can check that we always come back to the original quiver gauge theory after doing finite number of Seiberg dualities!

\begin{table}
\begin{center}
\begin{tabular}{| l | c | c|c|c| }   
  \hline                        
   ~&$(N_1,N_2)$ & $(k_1^{-},k_2^{-})$ & $(k_1^{+},k_2^{+})$  \\ \hline
    $\mu_1(\text{initial})$ &$(2,2)$&$(3,-3)$&$(1,-1)$\\ \hline
   $\mu_2$&$(3,2)$&$(-3,-3)$&$(-1,0)$\\ \hline
     $\mu_1$&$(3,5)$&$(-6,3)$&$(-1,0)$\\ \hline
   $\mu_2$&$(3,5)$&$(6,-3)$&$(1,0)$\\ \hline
   $\mu_1$&$(3,2)$&$(3,3)$&$(1,0)$\\ \hline
\text{final}&$(2,2)$&$(-3,3)$&$(-1,1)$\\ \hline
  \hline  
\end{tabular}
\end{center}
  \caption{The change of gauge group ranks and CS levels under sequences of Seiberg duality on $A_2$ quiver.}
    \label{data}
\end{table}

\newpage
\subsubsection{Mutation finite theories}
A second question about classification one might ask is: what is the shape of quiver such that there are only finite number of quivers in doing dualities (the choice of rank and CS level can be infinite). This is 
also answered in cluster algebra literature, and almost all of them comes from the quiver derived from triangulated surface \cite{fomin-2008-201}.  Basically, one can first triangulate the bordered Riemann surface
and then find a quiver from triangulation.

\begin{center}
\begin{figure}[htbp]
\small
\centering
\includegraphics[width=16cm]{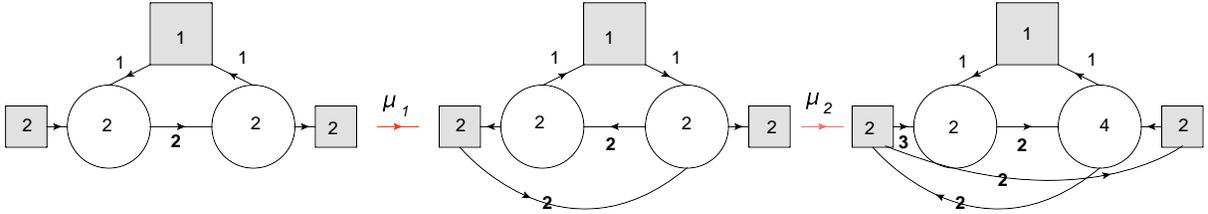}
\caption{A CSM quiver gauge theory based on affine $A_1$  Dynkin diagram, we use number on the quiver arrows to denote the multiple quiver arrows. 
Here we draw the extended quiver $\tilde{Q}$. We show quiver and gauge group ranks for several dualities, and 
there are infinite number of duality frames.}
\label{affine}
\end{figure}
\end{center}

\textbf{Example} The simplest example is a quiver based on affine $A_1$ Dynkin diagram, see figure. \ref{affine}. 
Let's choose the rank of gauge group and flavor group as $(s_1, N_1, N_2, s_2)=(2,2,2,2)$. We 
take the bare CS level of two dynamical nodes as $(k_1, k_2)=(0, 0)$.
We can do a sequence of Seiberg duality and find that 
there are only one type of quiver diagram (dynamical part), but the gauge group rank and CS levels can take infinite set of values!

\subsubsection{Network on Riemann surface}
All the other quiver gauge theory might be called wild type as  the behavior under dualities is quite wild. However, if we confine ourselves to certain subset of dualities, the theory under duality 
is actually under control. 

There is a huge class of nice  quiver gauge theories which can be read from the bipartite network on bordered Riemann surface. In the case of bipartite network on torus,  a lot of studies has been done
in \cite{Franco:2005rj}, and see \cite{Xie:2012mr,Franco:2012mm} for the generalization to other bordered Riemann surface for the 4d ${\cal N}=1$ context. 
Quiver gauge theories from network have nice behavior under duality if we only perform duality on  quiver nodes with four quiver arrows. This type of special dualities  can be represented by  
a geometric operation on network called square move. Regarding the quiver from network as 3d theory, and only do above special dualities,
we still have good control about these theories such as classification, etc.

\newpage

\subsection{Periodicity of Seiberg duality}
We have shown that the gauge theory comes back to itself after doing five dualities for the $A_2$ quiver. This phenomenon is quite common for a large class of quivers, and it plays an essential role in studying
wall crossing phenomenon of four dimensional $\mathcal{N}=2$ theory. It would be interesting to see whether this periodicity property of Seiberg duality have any applications to 
gauge theory dynamics.

\subsection{Positivity and SUSY breaking}
We have not talked about anything on the existence of the superconformal field theory for an arbitrary quiver. 
In general it is a difficult question to answer. However, there is an obvious criteria from duality point of view:

\textbf{Conjecture}: SUSY is broken if gauge group rank is negative in one of duality frame.

 When there is only one  $\text{U}(N_c)$ group  with $N_f$ non-chiral matter, it is known that 
 supersymmetry is  broken for  $N_f<N_c$, and the gauge group of dual theory indeed has negative ranks. 
In general, it is quite difficult to know whether an initial choice of gauge group ranks and Chern-Simons levels would ensure that SUSY is unbroken, as typically the quiver is of 
infinite type. 

The question about positivity of gauge group ranks  is actually closely related to the positivity property of tropical cluster algebra. If we consider quiver gauge theory  from triangulated 
Riemann surface, it is possible to prove that certain initial choice of gauge groups will ensure that the gauge groups are always positive in any other duality frame. We do not 
discuss details about these models, and leave the study of this interesting class to the future work.

\subsection{Moduli space and monopole operators}
The scalar potential of a general YM-CS quiver gauge theory is
\begin{align}
V=\sum{e_i^2\over 2}(D_i-\zeta_i^{eff})^2+\sum_{\alpha:i\rightarrow j}(\sigma_iX_{\alpha}-X_{\alpha}\sigma_j)^2+\sum_{\alpha:i\rightarrow j} ({\partial W\over \partial X_{\alpha}})^2
\end{align}
here $\sigma$ is the real scalar in vector multiplet, and $D_i$ is the familiar 4d D-terms:
\begin{equation}
D_i=\sum_{\alpha:i\rightarrow j} X_{\alpha}X_{\alpha}^{+}-\sum_{\beta:j\rightarrow i} X_{\beta}^{+}X_{\beta}
\end{equation}
with plus (minus) sign for fundamental (anti-fundamental) matter charged under $i$th gauge group. The vanishing of the potential leads to
\begin{equation}
D_i=\zeta_i^{eff},~~~\sigma_iX_{\alpha}-X_{\alpha}\sigma_j=0,~~{\partial W\over \partial X_{\alpha}}=0.
\label{moduli}
\end{equation}
$\sigma_i$ can be brought into diagonal form $\sigma_i=\text{diag}(\sigma_1^{(i)},\ldots,\sigma_{n_i}^{(i)},\ldots,\sigma_{N_i}^{(i)})$ using gauge symmetry, and the effective
FI term is 
\begin{equation}
\zeta^{eff}_{n_i}=\zeta_i+k_i\sigma_{n_i}^{(i)}+{1\over 2}\sum_{j=1}\sum_{n_j=1}^{N_j}\epsilon_{ij}|\sigma_{n_i}^{(i)}-\sigma_{n_j}^{(j)}|,
\label{FI}
\end{equation}
here $\zeta_i$ is the bare FI term and the real mass terms of the flavor group are also denoted by $\sigma$. 
In the following, we are going to consider some combinatorial aspects of the moduli space under the duality.

\textbf{Geometric branch}:
Consider a quiver without any flavor groups.
There is a geometric branch by setting $\sigma_i=\sigma I_i$ with $I_i$ the identity matrix, and the second equation of [\ref{moduli}] is automatically satisfied. The remaining two vacuum equations read
\begin{equation}
D_i=\sigma k_i  I_i,~~~{\partial W\over \partial X_{\alpha}}=0,
\end{equation}
so at fixed $\sigma$, the above equation is exactly the same as the vacuum equations of 4d theory with the CS levels play the role of FI term. 
There are two differences: first $\sigma$ is now also a coordinate of 
the moduli space, and second there is a overall gauge field whose dual give another real flat direction, and therefore the 3d moduli space has one more complex dimension. 

Since there is no flavor groups, then to have generic solution, we must have $\sum_i k_i N_i=0$ which is derived by
taking  the trace of $D$ term equation. However, we now claim that this condition is not duality invariant!
After doing Seiberg duality on quiver node $i$, using the duality formula we have 
\begin{equation}
\begin{cases}
\sum_i k_i^{'} N_i^{'}=\sum_i k_i N_i=0,~~~~~~~~~~~~~~~~~~~~~~~~|k_i|\leq {1\over 2}|s_1-s_2| \\
\sum_i k_i^{'} N_i^{'}=\sum_i k_i N_i-(k_i^{2}-{1\over4}(s_1-s_2)^2),~~ |k_i|>{1\over 2}|s_1-s_2|\\
\end{cases}
\end{equation}
so the condition $\sum_i k_i N_i=0$ is not always duality invariant! 

\textbf{Monopole operator}:
Instead of looking at the scalar potential, one can describe the moduli space using the gauge invariant operators and the relation between them, i.e. the chiral ring.
For four dimensional quiver gauge theory, gauge invariant operator is related to the oriented loops in the quiver, and one can deal with these loops using path algebra (when
there are superpotential term, we actually deal with Jacobian algebra).

For 3d quiver gauge theory, we still have the gauge invariant operators associated with the loops in the quiver $Q$. We now have extra monopole operators $T$ and $\tilde{T}$.
We would like to consider the simplest diagonal monopole operator, which will create a magnetic flux
\begin{equation}
T:~~~(1,0,\ldots,0),~~~~~~\tilde{T}:~~~(-1,0,\ldots,0)
\end{equation}
In the presence of the CS term and the chiral matter, the above monopole operators will acquire following electric charge \cite{Benini:2011cma}:
\begin{equation}
q_i(T)=k_i^{-},~~~q_i(\tilde{T})=-k_i^{+}.
\end{equation}
It is curious to notice that this number is  the same as the number of new arrows between $i$th node and extra node. So the extra node 
in quiver $\tilde{Q}$ would be thought of as the monopole operators creating diagonal magnetic flux for every gauge groups. Similarly, 
the extra node in $\tilde{Q}^o$ can be thought of as diagonal monopole operators $T$. One can read the electric charges of these monopole operators
in any duality frame from the quiver diagram $\tilde{Q}$ and $\tilde{Q}^o$: they are simply the number of arrows between original quiver nodes and the extra node. 
With this geometric realization of monopole operators, it is natural to expect that the gauge invariant operators involving monopole operators can be read from  extended quiver, and 
it would be interesting to further study this problem.

\section{Conclusion}
In this paper, We show that 3d Seiberg-like duality of a CSM  quiver gauge theory $Q$ is identified with tropical cluster algebra
of an extended quiver $\tilde{Q}$: simply adding an extra quiver node with rank one and adding  arrows  
determined by CS levels.  We then use cluster algebra results to study some global aspects of dualities. 

We focus on combinatorial aspects of 3d $\mathcal{N}=2$ Seiberg duality. It is interesting 
to further study the dynamical properties of these theories, like Witten index, moduli space, partition function, index, etc, and we hope
duality would play an important role.

One has very similar Seiberg duality on the same quiver if we regard it as defining a  two dimensional $(2,2)$ theory or one dimensional $\mathcal{N}=4$ theory.
The similar behavior of Seiberg duality of theory in different space-time dimensions can be seen from the brane construction \cite{Giveon:1998sr}: the Seiberg duality can be roughly 
seen as moving the relative position of different branes, which can be implemented on the brane systems engineering quiver gauge theories in different space-time dimension. 
Although Seiberg duality for quiver gauge theory is quite similar in two and one dimensions, there are some very interesting differences, and details will be presented  elsewhere. 

There is a different kind of IR equivalence namely the mirror symmetry for 3d $\mathcal{N}=2$ theory \cite{Aharony:1997bx,deBoer:1997ka}. Mirror symmetry matches the Coulomb branch of original theory to 
the Higgs branch of mirror theory, and the mirror theory typically looks 
very different from original theory (sometimes it maps a non-Lagrangian theory to a Lagrangian theory). Moreover, it seems that mirror symmetry can not be derived using local operations on gauge groups like Seiberg duality. In this respect, it is kind of difficult to find mirror theory. 
Brane construction helps a lot in finding mirrors for 3d $\mathcal{N}=4$ theory \cite{Hanany:1996ie}. The study on mirror symmetry for non-abelian 3d $\mathcal{N}=2$ gauge theory is relatively few, 
and it would be interesting to find more non-abelian mirror pairs. For the particular three dimensional  $\mathcal{N}=2$ theory derived from M5 branes on a Riemann surface times a circle \cite{Xie:2013gma}, it seems quite possible 
to find their  mirrors following \cite{Benini:2010uu}.

\begin{flushleft}
\textbf{Acknowledgments}
\end{flushleft}
We would like to thank F. Benini, Peng Zhao for many helpful discussions. 
This research is supported in part by Zurich Financial services membership and by the U.S. Department of Energy, grant DE-SC0009988. 

\bibliographystyle{utphys} 
 \bibliography{PLforRS}    

\providecommand{\href}[2]{#2}\begingroup\raggedright\begin{thebibliography}{10}

\bibitem{Seiberg:1994pq}
N.~Seiberg, ``{Electric - magnetic duality in supersymmetric nonAbelian gauge
  theories},'' \href{http://dx.doi.org/10.1016/0550-3213(94)00023-8}{{\em
  Nucl.Phys.} {\bf B435} (1995)  129--146},
\href{http://arxiv.org/abs/hep-th/9411149}{{\tt arXiv:hep-th/9411149
  [hep-th]}}.
%%CITATION = HEP-TH/9411149;%%.

\bibitem{Berenstein:2002fi}
D.~Berenstein and M.~R. Douglas, ``{Seiberg duality for quiver gauge
  theories},''
\href{http://arxiv.org/abs/hep-th/0207027}{{\tt arXiv:hep-th/0207027
  [hep-th]}}.
%%CITATION = HEP-TH/0207027;%%.

\bibitem{Fomin2001}
S.~{Fomin} and A.~{Zelevinsky}, ``{Cluster algebras I: Foundations},'' {\em J.
  Amer. Math. Soc.} {\bf 15} (2002)  497--529,
  \href{http://arxiv.org/abs/arXiv:math/0104151}{{\tt arXiv:math/0104151}}.

\bibitem{Xie:2012mr}
D.~Xie and M.~Yamazaki, ``{Network and Seiberg Duality},''
\href{http://arxiv.org/abs/1207.0811}{{\tt arXiv:1207.0811 [hep-th]}}.
%%CITATION = ARXIV:1207.0811;%%.

\bibitem{Franco:2012mm}
S.~Franco, ``{Bipartite Field Theories: from D-Brane Probes to Scattering
  Amplitudes},''
\href{http://arxiv.org/abs/1207.0807}{{\tt arXiv:1207.0807 [hep-th]}}.
%%CITATION = ARXIV:1207.0807;%%.

\bibitem{Aharony:1997gp}
O.~Aharony, ``{IR duality in d = 3 N=2 supersymmetric USp(2N(c)) and U(N(c))
  gauge theories},''
  \href{http://dx.doi.org/10.1016/S0370-2693(97)00530-3}{{\em Phys.Lett.} {\bf
  B404} (1997)  71--76},
\href{http://arxiv.org/abs/hep-th/9703215}{{\tt arXiv:hep-th/9703215
  [hep-th]}}.
%%CITATION = HEP-TH/9703215;%%.

\bibitem{Aharony:2013dha}
O.~Aharony, S.~S. Razamat, N.~Seiberg, and B.~Willett, ``{3d dualities from 4d
  dualities},'' \href{http://dx.doi.org/10.1007/JHEP07(2013)149}{{\em JHEP}
  {\bf 1307} (2013)  149},
\href{http://arxiv.org/abs/1305.3924}{{\tt arXiv:1305.3924 [hep-th]}}.
%%CITATION = ARXIV:1305.3924;%%.

\bibitem{Giveon:2008zn}
A.~Giveon and D.~Kutasov, ``{Seiberg Duality in Chern-Simons Theory},''
  \href{http://dx.doi.org/10.1016/j.nuclphysb.2008.09.045}{{\em Nucl.Phys.}
  {\bf B812} (2009)  1--11},
\href{http://arxiv.org/abs/0808.0360}{{\tt arXiv:0808.0360 [hep-th]}}.
%%CITATION = ARXIV:0808.0360;%%.

\bibitem{Benini:2011mf}
F.~Benini, C.~Closset, and S.~Cremonesi, ``{Comments on 3d Seiberg-like
  dualities},'' \href{http://dx.doi.org/10.1007/JHEP10(2011)075}{{\em JHEP}
  {\bf 1110} (2011)  075},
\href{http://arxiv.org/abs/1108.5373}{{\tt arXiv:1108.5373 [hep-th]}}.
%%CITATION = ARXIV:1108.5373;%%.

\bibitem{Amariti:2009rb}
A.~Amariti, D.~Forcella, L.~Girardello, and A.~Mariotti, ``{3D Seiberg-like
  Dualities and M2 Branes},''
  \href{http://dx.doi.org/10.1007/JHEP05(2010)025}{{\em JHEP} {\bf 1005} (2010)
   025},
\href{http://arxiv.org/abs/0903.3222}{{\tt arXiv:0903.3222 [hep-th]}}.
%%CITATION = ARXIV:0903.3222;%%.

\bibitem{Cremonesi:2010ae}
S.~Cremonesi, ``{Type IIB construction of flavoured ABJ(M) and fractional M2
  branes},'' \href{http://dx.doi.org/10.1007/JHEP01(2011)076}{{\em JHEP} {\bf
  1101} (2011)  076},
\href{http://arxiv.org/abs/1007.4562}{{\tt arXiv:1007.4562 [hep-th]}}.
%%CITATION = ARXIV:1007.4562;%%.

\bibitem{Xie:2012gd}
D.~Xie, ``{BPS spectrum, wall crossing and quantum dilogarithm identity},''
\href{http://arxiv.org/abs/1211.7071}{{\tt arXiv:1211.7071 [hep-th]}}.
%%CITATION = ARXIV:1211.7071;%%.

\bibitem{Gaiotto:2010be}
D.~Gaiotto, G.~W. Moore, and A.~Neitzke, ``{Framed BPS states},''
  \href{http://arxiv.org/abs/1006.0146}{{\tt arXiv:1006.0146 [hep-th]}}.

\bibitem{Alim:2011kw}
M.~Alim, S.~Cecotti, C.~Cordova, S.~Espahbodi, A.~Rastogi, and C.~Vafa, ``{N=2
  Quantum Field Theories and Their BPS Quivers},''
\href{http://arxiv.org/abs/1112.3984}{{\tt arXiv:1112.3984 [hep-th]}}.
%%CITATION = 1112.3984;%%.

\bibitem{Xie:2013lca}
D.~Xie, ``{Higher laminations, webs and N=2 line operators},''
\href{http://arxiv.org/abs/1304.2390}{{\tt arXiv:1304.2390 [hep-th]}}.
%%CITATION = ARXIV:1304.2390;%%.

\bibitem{arkani2012scattering}
N.~Arkani-Hamed, J.~L. Bourjaily, F.~Cachazo, A.~B. Goncharov, A.~Postnikov,
  and J.~Trnka, ``Scattering amplitudes and the positive grassmannian,'' {\em
  arXiv preprint arXiv:1212.5605} (2012)  .

\bibitem{Dimofte:2013iv}
T.~Dimofte, M.~Gabella, and A.~B. Goncharov, ``{K-Decompositions and 3d Gauge
  Theories},''
\href{http://arxiv.org/abs/1301.0192}{{\tt arXiv:1301.0192 [hep-th]}}.
%%CITATION = ARXIV:1301.0192;%%.

\bibitem{Terashima:2013fg}
Y.~Terashima and M.~Yamazaki, ``{3d N=2 Theories from Cluster Algebras},''
\href{http://arxiv.org/abs/1301.5902}{{\tt arXiv:1301.5902 [hep-th]}}.
%%CITATION = ARXIV:1301.5902;%%.

\bibitem{fock2005dual}
V.~V. Fock and A.~B. Goncharov, ``Dual teichmuller and lamination spaces,''
  {\em arXiv preprint math/0510312} (2005)  .

\bibitem{Niarchos:2008jb}
V.~Niarchos, ``{Seiberg Duality in Chern-Simons Theories with Fundamental and
  Adjoint Matter},''
  \href{http://dx.doi.org/10.1088/1126-6708/2008/11/001}{{\em JHEP} {\bf 0811}
  (2008)  001},
\href{http://arxiv.org/abs/0808.2771}{{\tt arXiv:0808.2771 [hep-th]}}.
%%CITATION = ARXIV:0808.2771;%%.

\bibitem{Kim:2013cma}
H.~Kim and J.~Park, ``{Aharony Dualities for 3d Theories with Adjoint
  Matter},'' \href{http://dx.doi.org/10.1007/JHEP06(2013)106}{{\em JHEP} {\bf
  1306} (2013)  106},
\href{http://arxiv.org/abs/1302.3645}{{\tt arXiv:1302.3645 [hep-th]}}.
%%CITATION = ARXIV:1302.3645;%%.

\bibitem{Closset:2012eq}
C.~Closset, ``{Seiberg duality for Chern-Simons quivers and D-brane
  mutations},'' \href{http://dx.doi.org/10.1007/JHEP03(2012)056}{{\em JHEP}
  {\bf 1203} (2012)  056},
\href{http://arxiv.org/abs/1201.2432}{{\tt arXiv:1201.2432 [hep-th]}}.
%%CITATION = ARXIV:1201.2432;%%.

\bibitem{Aharony:1997bx}
O.~Aharony, A.~Hanany, K.~A. Intriligator, N.~Seiberg, and M.~Strassler,
  ``{Aspects of N=2 supersymmetric gauge theories in three-dimensions},''
  \href{http://dx.doi.org/10.1016/S0550-3213(97)00323-4}{{\em Nucl.Phys.} {\bf
  B499} (1997)  67--99},
\href{http://arxiv.org/abs/hep-th/9703110}{{\tt arXiv:hep-th/9703110
  [hep-th]}}.
%%CITATION = HEP-TH/9703110;%%.

\bibitem{Intriligator:2013lca}
K.~Intriligator and N.~Seiberg, ``{Aspects of 3d N=2 Chern-Simons-Matter
  Theories},''
\href{http://arxiv.org/abs/1305.1633}{{\tt arXiv:1305.1633 [hep-th]}}.
%%CITATION = ARXIV:1305.1633;%%.

\bibitem{Fock:math0311245}
V.~V. Fock and A.~B. Goncharov, ``Cluster ensembles, quantization and the
  dilogarithm,'' \href{http://arxiv.org/abs/math/0311245}{{\tt math/0311245}}.

\bibitem{fomin2003cluster}
S.~Fomin and A.~Zelevinsky, ``Cluster algebras ii: Finite type
  classification,'' {\em Inventiones Mathematicae} {\bf 154} (2003) no.~1,
  63--121.

\bibitem{fomin-2008-201}
S.~Fomin, M.~Shapiro, and D.~Thurston, ``Cluster algebras and triangulated
  surfaces. part i: Cluster complexes,'' {\em ACTA MATHEMATICA} {\bf 201}
  (2008)  83. \url{arXiv.org:math/0608367}.

\bibitem{Franco:2005rj}
S.~Franco, A.~Hanany, K.~D. Kennaway, D.~Vegh, and B.~Wecht, ``{Brane Dimers
  and Quiver Gauge Theories},''
  \href{http://dx.doi.org/10.1088/1126-6708/2006/01/096}{{\em JHEP} {\bf 01}
  (2006)  096},
\href{http://arxiv.org/abs/hep-th/0504110}{{\tt arXiv:hep-th/0504110}}.
%%CITATION = HEP-TH/0504110;%%.

\bibitem{Benini:2011cma}
F.~Benini, C.~Closset, and S.~Cremonesi, ``{Quantum moduli space of
  Chern-Simons quivers, wrapped D6-branes and AdS4/CFT3},''
  \href{http://dx.doi.org/10.1007/JHEP09(2011)005}{{\em JHEP} {\bf 1109} (2011)
   005},
\href{http://arxiv.org/abs/1105.2299}{{\tt arXiv:1105.2299 [hep-th]}}.
%%CITATION = ARXIV:1105.2299;%%.

\bibitem{Giveon:1998sr}
A.~Giveon and D.~Kutasov, ``{Brane dynamics and gauge theory},''
  \href{http://dx.doi.org/10.1103/RevModPhys.71.983}{{\em Rev.Mod.Phys.} {\bf
  71} (1999)  983--1084},
\href{http://arxiv.org/abs/hep-th/9802067}{{\tt arXiv:hep-th/9802067
  [hep-th]}}.
%%CITATION = HEP-TH/9802067;%%.

\bibitem{deBoer:1997ka}
J.~de~Boer, K.~Hori, Y.~Oz, and Z.~Yin, ``{Branes and mirror symmetry in N=2
  supersymmetric gauge theories in three-dimensions},''
  \href{http://dx.doi.org/10.1016/S0550-3213(97)00444-6}{{\em Nucl.Phys.} {\bf
  B502} (1997)  107--124},
\href{http://arxiv.org/abs/hep-th/9702154}{{\tt arXiv:hep-th/9702154
  [hep-th]}}.
%%CITATION = HEP-TH/9702154;%%.

\bibitem{Hanany:1996ie}
A.~Hanany and E.~Witten, ``{Type IIB superstrings, BPS monopoles, and
  three-dimensional gauge dynamics},''
  \href{http://dx.doi.org/10.1016/S0550-3213(97)00157-0}{{\em Nucl. Phys.} {\bf
  B492} (1997)  152--190},
\href{http://arxiv.org/abs/hep-th/9611230}{{\tt arXiv:hep-th/9611230}}.
%%CITATION = HEP-TH/9611230;%%.

\bibitem{Xie:2013gma}
D.~Xie, ``{M5 brane and four dimensional N=1 theories I},''
\href{http://arxiv.org/abs/1307.5877}{{\tt arXiv:1307.5877 [hep-th]}}.
%%CITATION = ARXIV:1307.5877;%%.

\bibitem{Benini:2010uu}
F.~Benini, Y.~Tachikawa, and D.~Xie, ``{Mirrors of 3d Sicilian theories},''
  \href{http://dx.doi.org/10.1007/JHEP09(2010)063}{{\em JHEP} {\bf 1009} (2010)
   063}, \href{http://arxiv.org/abs/1007.0992}{{\tt arXiv:1007.0992 [hep-th]}}.

\end{thebibliography}\endgroup
\end{document}